# Multiple oligo nucleotide arrays: Methods to reduce manufacture time and cost

Kang Ning
University of Michigan, Ann Arbor
Ann, Arbor, MI, USA
kning@umich.edu

**Abstract**

The customized multiple arrays are becoming vastly used in microarray experiments for varies purposes, mainly for its ability to handle a large quantity of data and output high quality results. However, experimenters who use customized multiple arrays still face many problems, such as the cost and time to manufacture the masks, and the cost for production of the multiple arrays by costly machines. Although there is some research on the multiple arrays, there is little concern on the manufacture time and cost, which is actually important to experimenters.

In this paper, we have proposed methods to reduce the time and cost for the manufacture of the customized multiple arrays. We have first introduced a heuristic algorithm for the mask decomposition problem for multiple arrays. Then a streamline method is proposed for the integration of different steps of manufacture on a higher level. Experiments show that our methods are very effective in reduction of the time and cost of manufacture of multiple arrays.

## 1. Introduction

The array-based method is a method of choice for many biological sequence analysis experiments. The profiling of the microarray can provide much information for experimenters. The DNA microarrays are composed of a set of distinct nucleic acid samples arranged in a regular lattice of spots on a solid support. The hybridization of RNA and DNA-derived samples with fluorescent labels to DNA microarrays allows the monitoring of gene expression or polymorphisms in genomic DNA [1]. Currently, the two widely used formats of DNA chips are the cDNA arrays [2] and customized oligonucleotides arrays [1, 3, 4]. The focus of this paper is on the oligonucleotide arrays.

The oligonucleotide arrays are deposited onto the solid support using either int-jet method [5], photolithographic method [6] or maskless method [7]. In the photolithographic method (used by Affymetrix), a specific mask is fabricated for each cycle of nucleotide addition that permits light to penetrate only at positions where nucleotides are to be added. A synthesis cycle consists of shining light through the mask onto the chip surface. The positions where light passes through the mask and reaches the chip are activated for synthesis by the removal of a photolabile protective group from the exposed end of the oligonucleotide. By this means, the pattern in which light penetrates the masks directs the base by base synthesis of oligonucleotides on a solid surface. After photodeprotection the chip is washed in a solution containing a single nucleotide that binds to oligonucleotides at the deprotected positions. This method results in the in situ synthesis of oligonucleotides on an array surface. Light-directed chemical synthesis has been used to produce arrays with as many as several millions of DNA oligos (representing up to 9,000 genes) with minimal cross-hybridization or inter-feature variability [1]. The int-jet method is cheaper and more accurate for long oligos, and maskless method is both cheaper and faster, but they are not vastly used in industry (though int-jet method is widely used in academic world).

As the number of genes to be examined is becoming larger (say 100,000 genes), as well as the increasing number of repeated experiments which by itself requires multiple arrays, one oligonucleotide arrays is not enough to perform the experiments which require a few million oligonucleotides. For multiple arrays, each array can be made relatively small, and small arrays also have the advantage that their manufacture is easier, cheaper and faster, and the qualities controls are easier for experimenters. In practice, the multiple array systems are feasible, while more and more such systems are available, with the cost of microarray decrease, and the scale of microarray experiments become large [8].

There are some research on the synthesis of multiple arrays [9, 10], and it is noticed that the experimenters still faces some problems for the actual synthesis of the multiple arrays, but there is little research on the reduction of manufacture time and cost. The time and cost of manufacture multiple oligo arrays can be reduced by shortening the synthesis sequence for the oligos [9-11], and more profoundly by the better method to manufacture masks and more compact procedures on all oligo arrays. We will focus on better method to manufacture masks, as well as short process steps to syntehsis all oligo arrays.

For all of the synthesis methods, including the maskless method, the multiple arrays introduce the problem on how to make the deposition process parallel on different arrays, which is directly related with the time (and cost) to manufacture. On a higher level of synthesis of multiple arrays, there are mainly 3 steps: generation of deposition strategy, manufacture the masks and processing of oligos. For multiple arrays, all of these 3 steps would be accomplished by different machines, and each of them need a considerable time to process. Thus they can be streamlined (parallelized), so that the speed is further accelerated and cost further reduced. (The idea of streamline should be attributed to Henry Ford, who introduced mass production to the population.) The shorter time and less money needed by the new method are very important since that means the acceleration of the biological analysis by microarray experiments.

Since the maskless method is to some extend constrained by the size of the oligos array, and faces problem when larger arrays are used in microarray experiments. So the physical masks will still be of great use for researchers. For the synthesis methods which requires physical masks (for most of the current experiments), the fabrication of masks is both time consuming and costly. Each of the physical masks is fabricated in a process similar to that of block printing, and the hard effort put on this process make it time consuming and costly. Since masks are composed of many rectangles, we can also select a set of rectangles and fabricate the mask at each deposition step, thus can greatly reduce the time and money needed in the actual synthesis. By this way, we only need a number of rectangles (building blocks), and compose the (This idea should be attributed to Bi Sheng, who invented movable type printing over 1,000 years ago.) And these lead to the introduction of mask decomposition problem (MDP) (or rectangle cover problem) [12] for multiple arrays, for which the reduction of time and cost are valuable.

In this paper, we will explain our methods to reduce the time and cost for the manufacture of the multiple oligos arrays. We will concentrate on the mask decomposition problem (MDP) for multiple arrays, and the streamline process for the whole manufacture process.

2. Material and Methods

The reduction of time and cost of manufacture of multiple arrays is vital, but there are many problems. We have focused on two main problems for this issue: the mask decomposition problem for multiple arrays and the parallelization of the processes; and we have proposed methods to solve these problems.

**Mask Decomposition Problem for Multiple Arrays**

The mask decomposition problem (MDP) is introduced by Hannenhalli *et.al.* [12] for single oligos arrays. The solution for the mask decomposition problem is a set of the rectangles that can compose the masks. On multiple arrays, the problem becomes more important, since a good solution means a great reduction of cost in manufacture. However, since there are more masks for multiple arrays, the problem becomes more complicated. A sophisticated method for the mask decomposition problem may not give good results for the multiple arrays, because the resulting rectangles one array may not present multiple times in other arrays.

The mask decomposition problem for multiple arrays is an extension of the mask decomposition problem. Suppose we are given $p$ masks to be manufactured. We say that a mask is composed of some rectangles if these rectangles can be put together in certain manner to form the mask. The solution to this problem is a set of $q$ rectangles $r_1, r_2 ... r_q$, so that any of the $p$ masks is composed of a subset of the $p$ masks. An example of the mask decomposition problem is illustrated in Figure 1.

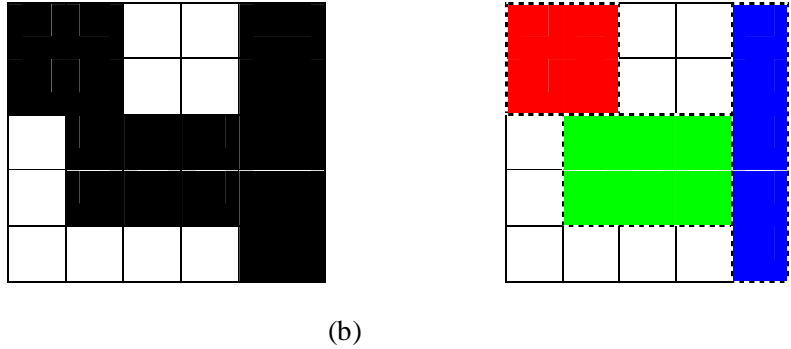

(a) (b)

**Figure 1. An illustration of the mask decomposition problem. In this example, the mask (a) is composed of 3 rectangles (b).**

We have assumed that the synthesis sequences are generated by the LAP algorithm [11]. The strategies for the generation of synthesis sequences are extensively investigated [13], and the LAP algorithm is fast and can produce short synthesis sequences. For the placement of the oligos on the array, we also assumed that the oligos are to be placed on the arrays row by row, from left to right. The placement of oligos is beneficial to the solution to the mask decomposition problem [12] for multiple arrays, but it is a rather complicated problem which is not our focus.

For the mask manufacture of multiple arrays, there is the strait forward method is to manufacture the masks anew one at a time, and use it for one cycle of oligos process. In general, each of the masks needs considerable time to manufacture; and since in most of the time it is used only once, this method is costly.

We have proposed a heuristic algorithm for this problem. In this algorithm, we have analyzed each of the masks in a row-by-row manner. For each of the row (or column), a continuous segment is selected as a rectangle. For masks of size $m*m$, there are at most $m$ shapes of rectangles ($1*1, 1*2 ... 1*m$) to compose all of the masks. Each of the rectangles has many copies, for example, for a $1*i$ rectangle, there are at most $\left\lfloor \dfrac{m}{i} \right\rfloor * m$ needed for one mask.

To compose a mask from rectangles, we can place the rectangles horizontally or vertically in the mask. In this paper, we have proposed a simple strategy to choose placements. For each of the spot to be protected, and is not examined, check the longest horizontal and vertical rectangles that cover this spot; the longest one is chosen, and all the spot covered by this rectangle are set to be examined. The importance of this mask decomposition method is that the rectangles generated can be reused to compose many masks for multiple arrays. The ease of manufacture many identical rectangles make this method practical in real experiments.

There is no guarantee that this method will result in less number of shapes than only using the horizontal rectangles; but it showed good results in experiments. A cartoon illustration of this method is illustrated in Figure 2. The procedure of this mask decomposition method is listed in Figure 3.

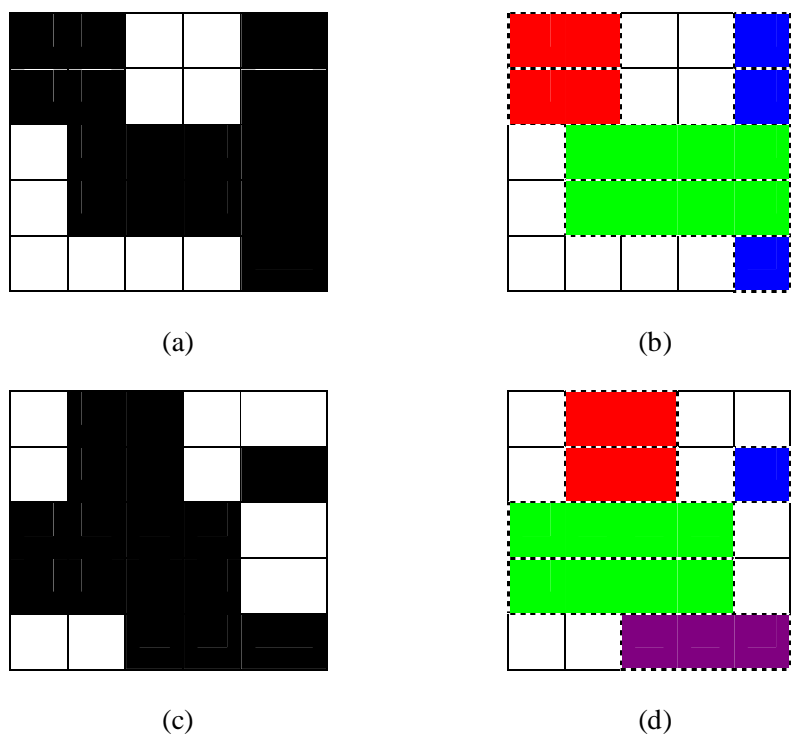

(a)  (b)

(c)  (d)

**Figure 2. An example of the method for mask decomposition problem for multiple arrays. Only horizontal rectangles are used. The two arrays (a) and (c) are decomposed to rectangles as in (b) and (d). There are 4 shapes of rectangles to compose the two masks.**

```
Mask Decomposition (O, S)

Input:     A set of arrays, A={O₁, O₂ … Oₙ},
           each array is composed of many oligos, Oᵢ={o₁, o₂ … oₘ*ₘ};
           A set of synthesis sequences, S={s₁, s₂ … sₙ};

A set of Hashs R={R₁, R₂ … Rₙ} is empty;
For each of the arrays Oᵢ and synthesis sequence sᵢ
      Generate p masks of size m*m, Mask={mask₁, mask₂ … maskₚ}, from Oᵢ
      and sᵢ, where p=length(sᵢ);
      For each of the masks
```

```
        For each of the spots
            If (this spot is to be masked) and (this spot is not
            examined)
                Search for longest horizontal rectangle 1*j
                that cover this spot;
                Search for longest vertical rectangle 1*k that
                cover this spot;
                r_i is rectangle of size 1*max(j,k);
                Push rectangle r_i into hash R_i, R_i(r_i)++;
                All spots covered by r_i are examined;
            EndIf
        EndFor
    EndFor
    Store hash R_i for O_i and s_i;
EndFor

Output: Hashes set R;
```

**Figure 3: the procedure of the mask decomposition method.**

**Streamlines**

The parallelization of the processes can be represented as this. Given $n$ steps for the synthesis of one array, with the time for each step $t_1, t_2 \ldots t_n$; and given k arrays to be synthesized. Find a function for the total synthesis time $T=f(t_1, t_2 \ldots t_n, k)$, so that T is as small as possible. For the synthesis of arrays, there are mainly 3 steps: $t_s$ for generation of deposition strategy, which generates the synthesis sequences by computers; $t_m$ for manufacture the masks, which mainly includes manufacturing of the masks; and $t_d$ for process of nucleic acids. For multiple arrays, all of these 3 steps can be accomplished by different machines, and each of them need a considerable time to process. Generally, the relative processing times are $t_s$ $t_m$ $t_d$.

For the parallelization of the processes, the strait forward method is to synthesis multiple arrays one by one, as illustrated in Figure 4. Assume that $t_s, t_m, t_p$ are the same for $k$ arrays, the total time needed is $k*(t_s+t_m+t_p)$.

| array 1 $t_s$ | array 1 $t_m$ | array 1 $t_d$ | array 2 $t_s$ | array 2 $t_m$ | array 2 $t_d$ | array 3 $t_s$ | … |

**Figure 4. The process of the strait forward method for the synthesis of multiple arrays.**

The parallelization of these steps is apparent. In Figure 5, we have illustrated the streamline process.

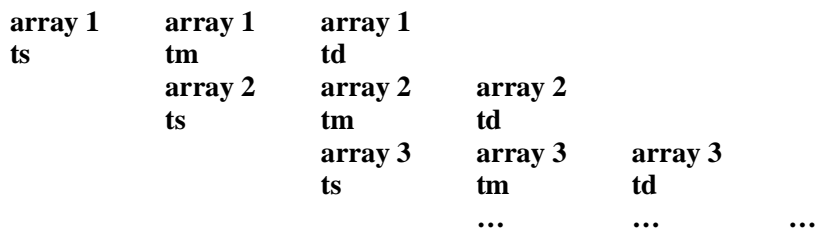

**Figure 5. The streamline process for the synthesis of the multiple arrays. Note that $t_s$, $t_m$ and $t_d$ are not of the same values, and different for different arrays.**

Since $t_s$ $t_m$ $t_d$ for general cases, and assume that $t_s, t_m, t_d$ are the same for $k$ masks, then the total time for the synthesis of multiple arrays is $t_s+t_m+k*t_d$. By strait forward method, $k*(t_s+t_m+t_d)$ time is needed. It is obvious that the more arrays to be synthesized; the more reduction can be made. This method is referred to as the "Simple Streamline" method.

Further parallelization is based on fine tuning of each steps. In each of the deposition step, we can parallelize these two steps to reduce the time further (about half of time can be reduced). For the deposition of nucleic acids, the physical deposition of nucleic acids is more time consuming, and most costly step, so the parallelization is concentrated on this step. We have devised a method that can reduce the time and cost significantly.

For multiple arrays, suppose their synthesis sequences are $\{s_1, s_2 ... s_n\}$, we find a short common supersequence $s^*$ for these synthesis sequences. The mask deprotection of each array is performed according to $s^*$, and all of the arrays which perform deprotection are subjected to one run of nucleic acids deposition. The possibly shortest common supersequence can be generated by LAP algorithm as described in [11], or by dynamic programming if there are few arrays ($n$ 4). By doing these, we can reduce the number of deposition steps to as few as possible. The number of process steps reduced is $s_1+s_2+...+s_n-s^*$. This procedure is illustrated in Figure 6.

```
Parallel deprotection and deposition (O, S)

Input:     A set of arrays, A={O₁, O₂ … On},
           each array is composed of many oligos, Oi={o₁, o₂ … om*m};
           A set of synthesis sequences, S={s₁, s₂ … sn};

Generate a possibly shortest common supersequence s* of S;
For every nucleic acids aaᵢ in s*
    For each of the array oᵢ with synthesis sequence sⱼ in S
        If sⱼ has aaᵢ at this stage
            Deprotect array oᵢ by specific mask;
        EndIf
    EndFor
    Deposit aaᵢ onto all of the arrays that performed the
    deprotection;
EndFor
```

**Figure 6. The procedure for parallelization of the deprotection and deposition steps.**

For example, suppose there are 4 arrays with synthesis sequences "ACG", "ACT", "CGT" and "AGT". Without this method, we will need 12 deposition steps. We can generate their shortest common supersequence "ACGT", and thus only 4 deposition steps are needed.

By using this method, both time and cost are reduced. It is obvious that the more arrays to be synthesized; the more reductions can be achieved by this way. The method using these additional techniques is called "Smart Streamline".

**Performance Ratio**

The reduction ratio $R_r$ can be represented as

$$R_r = \frac{T_p}{T_s}$$

Where $T_p$ is the time for the streamline process, and $T_s$ is the time by the strait forward method. For the synthesis of multiple arrays, the smaller the reduction ratio $R_r$, the better performance of the streamline method. For our mask decomposition method, assume that $t_s, t_m, t_d$ are the same for $k$ masks, $T_p = t_s + t_m + k*t_d$ and $T_s = (t_s + t_m + t_d)*k$, so

$$R_r = \frac{t_s + t_m + k*t_d}{(t_s + t_m + t_d)*k}$$

And this value approaches $t_d/(t_s+t_m+t_d)$ as the value of $k$ increase.

## 3. Experiments

We have applied our methods on many different datasets, including the simulated datasets and real oligos datasets. And we have compared the performance of our methods with the performance of strait forward methods.

**Datasets and experiment settings**

The experiments are performed on a PC with 3.0 GHz CPU and 1.0 GMb memory. The methods are implemented in Perl, and running on Linux OS.

The simulated oligos datasets are generated by random DNA oligos generator (http://www-personal.umich.edu/~kning/random.html). Short oligos are suitable for distinguishing between perfectly matched duplexes and single-base or two-base mismatches. Long oligo arrays [14] have better specificity and sensitivity for gene expression data, but the danger of non-specific cross-hybridization and manufacturing errors increases. In this paper, we have performed experiments on both short (25 bases) and long (30 to 50) simulated oligos arrays.

For multiple arrays of real oligos, we have obtained real gene sequences from GenBank [15] (ftp://ftp.ncbi.nih.gov/genbank/), and selected oligos by the Primer3 software [16]. The selected oligos are of 25 bases. These datasets are described in details in [9].

In experiments, we have used the LAP algorithm [11] to generate the short synthesis sequences. For each set of oligos on an array, the placement of the oligos follows the row-by-row, left-to-right order. We have first analyzed the performance of the strait forward methods. Then we have carefully performed our methods on the simulated and real arrays datasets, and compared the results with those from strait forward methods.

**Experiment procedure**

We have first analyzed the performance of the strait forward methods. For the manufacture of the masks, it was costly by strait forward method (cost more than $1,0000), and each mask needs about 45 minutes to synthesis.

For the strait forward method to synthesis multiple arrays one by one, the results were listed in Table 1. The time of $t_s$ was averaged out by 10 runs. We assumed that time of $t_m$ to be 45 minutes for each mask, which was the total time to synthesize the rectangles and compose all of the masks, divided by the number of masks. And we also assumed $t_d$ to be 30 minutes (5 minutes for deprotection, and 25 minutes for deposition) for each mask.

|  | Array size | No. of arrays | Time (minutes) | $T_s$ | $T_m$ | $T_d$ | Total | Simple Streamline | Smart Streamline |
|---|---|---|---|---|---|---|---|---|---|
| **Simulated** | | | | | | | | | |
| N5000K1000 | 100*100 | 3 | 79.7 | 40.18 | 3586.5 | 2391.0 | 18053.0 | 10799.7 | 5020.4 |
|  | 200*200 | 1 | 84.0 | 45.32 | 3780.0 | 2520.0 | 6345.3 | 6345.3 | 6326.3 |
| N10000K500 | 100*100 | 10 | 80.3 | 41.23 | 3613.5 | 2409.0 | 60637.3 | 27744.7 | 4257.4 |
|  | 200*200 | 3 | 83.0 | 46.65 | 3735.0 | 2490.0 | 18815.0 | 11251.7 | 5224.2 |
| N10000K1000 | 100*100 | 8 | 67.8 | 43.43 | 3051.0 | 2034.0 | 41027.4 | 19366.4 | 3731.2 |
|  | 200*200 | 2 | 82.0 | 50.78 | 3690.0 | 2460.0 | 12401.6 | 8660.8 | 5632.3 |
| N20000K500 | 100*100 | 10 | 81.4 | 45.63 | 3663.0 | 2442.0 | 61506.3 | 28128.6 | 4412.9 |
|  | 200*200 | 3 | 83.0 | 52.27 | 3735.0 | 2490.0 | 18831.8 | 11257.3 | 5212.8 |
| **Real set** | | | | | | | | | |
| gbest1 | 100*100 | 8 | 79.3 | 41.28 | 3568.5 | 2379.0 | 47910.2 | 22641.8 | 4342.0 |
| gbest1 | 200*200 | 2 | 76.3 | 49.53 | 3433.5 | 2289.0 | 11544.1 | 8061.0 | 5245.2 |
| gbest1 | 100*100 | 24 | 80.5 | 51.54 | 3622.5 | 2415.0 | 146137.0 | 61634.0 | 4189.3 |
| gbest1 | 200*200 | 6 | 81.5 | 57.91 | 3667.5 | 2445.0 | 37022.5 | 18395.4 | 4532.3 |

**Table 1. The time for each step of the synthesis process.**

From Table 1, we observed that for the simulated datasets that we have examined, the whole process needed about 100 to 480 hours (4 to 20 days) to complete, which were very long time for experimenters. For real datasets, the whole process needed about 150 to 1000 hours to complete, which are longer than on simulated datasets.

For the performance of the method for the mask decomposition problem for multiple arrays, we have applied our method to the simulated and real datasets. In the first test, we have only used the horizontal rectangles, and in the second test, we have used both the horizontal and vertical rectangles as we have explained. The results were shown in Table 2. For multiple arrays, the number of masks, number of shapes and other relevant numbers were average values on the number of arrays. The array sizes were chosen as 100*100 and 200*200, which were similar in real experiments.

| Datasets |  |  |  | Horizontal rectangles | | | Horizontal and Vertical rectangles | | |
|---|---|---|---|---|---|---|---|---|---|
|  | Array size | No. of arrays | No. of masks | No. of shapes | No. of rectangles | Rectangles per mask | No. of shapes | No. of rectangles | Rectangles per mask |
| **Simulated** | | | | | | | | | |
| N5000K1000 | 100*100 | 3 | 79.7 | 97.0 | 381420 | 4787.7 | 90.7 | 361155 | 4533.3 |
|  | 200*200 | 1 | 84.0 | 192.0 | 378538 | 4506.4 | 134.0 | 361069 | 4298.4 |
| N10000K500 | 100*100 | 10 | 80.3 | 99.5 | 1540131 | 19179.7 | 99.9 | 1452726 | 18091.2 |
|  | 200*200 | 3 | 83.0 | 195.7 | 1526149 | 18387.3 | 177.3 | 1448072 | 17446.7 |
| N10000K1000 | 100*100 | 8 | 67.8 | 83.0 | 766958 | 11306.5 | 83.3 | 723749 | 10669.5 |
|  | 200*200 | 2 | 82.0 | 187.0 | 758329 | 9247.9 | 163.5 | 720901 | 8791.5 |
| N20000K500 | 100*100 | 10 | 81.4 | 99.3 | 1541883 | 18942.1 | 99.6 | 2897333 | 36536.4 |
|  | 200*200 | 3 | 83.0 | 195.3 | 1524512 | 18367.6 | 177.7 | 1446970 | 17433.4 |
| **Real set** | | | | | | | | | |
| gbest1 | 100*100 | 8 | 79.3 | 99.0 | 1342628 | 16923.9 | 97.6 | 1265018 | 15945.6 |
|  | 200*200 | 2 | 76.3 | 176.7 | 1324909 | 17356.9 | 159.7 | 1253251 | 16418.1 |
| gbpln1 | 100*100 | 24 | 80.5 | 99.2 | 3481077 | 43231.5 | 98.9 | 3280712 | 40743.2 |
|  | 200*200 | 6 | 81.5 | 196.5 | 3434846 | 42145.4 | 190.2 | 3246212 | 39830.8 |

**Table 2. The statistical results about the mask decomposition process. Different rectangle placement settings were used. In simulated datasets, N was the number of genes, and K was the length of each gene.**

The results showed that if only horizontal rectangles were used, then for $m*m$ arrays, less than $m$ shapes were needed to compose masks. There were fewer shapes needed if both horizontal and vertical rectangles were used, and also fewer (more than 5% less) rectangles per mask than using only horizontal rectangles. For real datasets, these reductions were more obvious than for simulated datasets. The small number of shapes needed, as well as the few thousands of rectangles needed for each mask made it convenient to manufacture masks by this method.

We have also examined the distribution of the number of $1*i$ rectangles needed by the synthesis of one array. Parts of the results are shown in Figure 7.

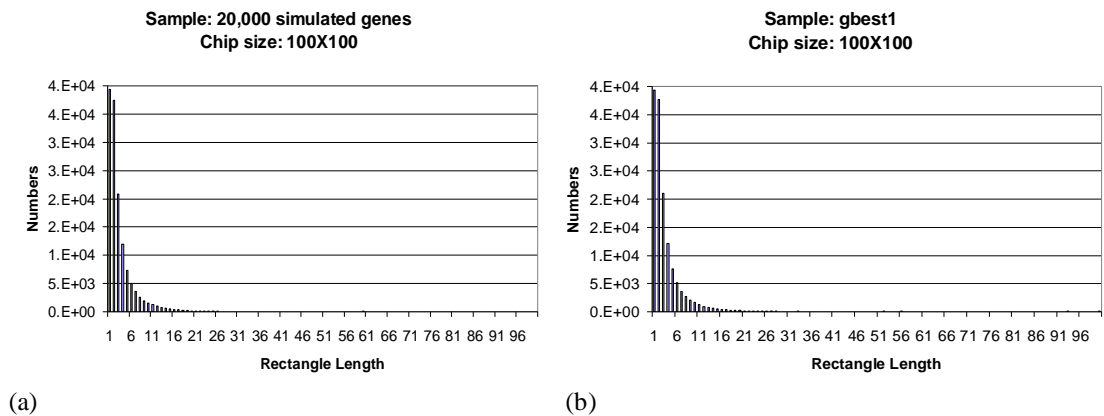

(a)                                                              (b)

**Figure 7: The distribution of the number of rectangles needed for the synthesis of one array. Both the horizontal and vertical rectangles were used. The results were based on (a) one array for simulated datasets and (a) one array for real datasets.**

From Figure 7, it was obvious that most of the $1*i$ rectangles had the length of 25 ( 100 for each $i$ 25), and more than 10,000 rectangles were needed for some of them ($i$=1, 2, 3, 4). For those rectangles with length $i$ 25, around 5 to 50 were needed to be synthesis for one array. We have also observed that (results not shown) these distributions were prevalent for multiple arrays, with similar quantities of rectangles for each array. Therefore, most of these rectangles could be reused for multiple arrays. Since most of the rectangles had small length, which were easier to make and can be used many time, our method was proved to be effective.

**Computational efficiency**

As for the computational time, the process of mask decomposition on one set of multiple arrays needed about 5 minutes. The time for the manufacture for all of the rectangles needed by multiple arrays was about 1 day. We assumed that the composition of masks from rectangles was 30 minutes for each mask. Then the manufacture of each of the mask was about 1day/$n$+30 minutes, where $n$ was the length of synthesis sequence. This was about 45 minutes for arrays with 25-mer oligos. Therefore, there was a great reduction from 1 day per mask in the strait forward method.

For the streamline method, the total synthesis times for both "Simple Streamline" and "Smart Streamline" were also listed in Table 1. As compared with the strait forward time, the reduction was significant. The "Simple Streamline" method could reduce the time needed by half. The performance ratios $R_r$ was 0.6, which meant we could use half of the time to synthesis all of the multiple arrays using streamline method.

For the "Smart Streamline" method, by parallelization of physical deprotection and physical deposition steps, we could further reduce the time needed. From Table 1, we have seen that the time was further reduced to about 50% to 5% of that needed by strait forward method. The number of the physical deposition was also reduced to about 100% to 10% of that needed by strait forward algorithm, depending on the number of arrays to be synthesized. Based on the streamline principle, the more arrays, the better performance ratios could be achieved by streamline method.

4.  Conclusions

In this paper, we have proposed some methods to effectively reduce the time and cost for manufacturing of the multiple oligos arrays. The methods for the mask decomposition problem for multiple arrays can greatly reduce the time and money needed to manufacture the masks. The streamline method for the whole synthesis processes can profoundly reduce the time needed, and with great scalability.

Experiments are encouraging. The mask decomposition method for multiple arrays can effective generates rectangles for the masks, with a few number of shapes for rectangles. For multiple arrays, the streamline method is very effective, with small performance ratio. Especially, the use of "Smart Streamline" method can greatly reduce the time and cost. The use of these methods can greatly reduce the time and money needed to synthesis multiple arrays. We believe that these methods can greatly benefit the microarray experiments based on multiple arrays, especially for researchers in academics.

The multiple arrays are becoming more and more popular now, and the multiple arrays introduce many problems. In this paper, we have put effort on the methods for reduction of manufacture time and cost, but our heuristic methods can be improved up on, and we are currently improving the methods. One of the possible improvements is that amino acids can be physically deposited onto multiple arrays in parallel, which can greatly reduce the cost. Among many problems introduced, we have only addressed a few of them, and there are still many other problems remain unsolved, such as the oligos placement problem and border minimization problem [12]. These interesting problems are also the targets of our further research.

**Acknowledgement**

We thank anonymous reviewers for their insightful comments.